\documentclass[prl,twocolumn,showpacs,preprintnumbers,amsmath,amssymb,floatfix]{revtex4}

\usepackage{graphicx}
\usepackage{dcolumn}
\usepackage{bm}

\begin{document}
\title{Networks and Cities: An Information Perspective}

\author{M. Rosvall$^{1,2}$}\email{rosvall@tp.umu.se}
\author{A. Trusina$^{1,2}$}
\author{P. Minnhagen$^{1,2}$}
\author{K. Sneppen$^{2}$}

\affiliation{$^{1)}$Department of Theoretical Physics, Ume{\aa} University,
901 87 Ume{\aa}, Sweden\\
$^{2)}$NORDITA, Blegdamsvej 17, Dk 2100, Copenhagen, Denmark}
\homepage{www.nordita.dk/research/complex}

\date{\today}

\begin{abstract}
Traffic is constrained by the information involved 
in locating the receiver and the
physical distance between sender and receiver.
We here focus on the former, and investigate traffic
in the perspective of information handling.
We re-plot the road map of cities in terms of 
the information needed to locate specific addresses and
create information city networks
with roads mapped to nodes and intersections to
links between nodes. These networks
have the broad degree distribution
found in many other complex networks.
The mapping to an information city network makes it 
possible to quantify the information associated with
locating specific addresses.
\end{abstract}

\pacs{89.70.+c,89.75.Fb,89.65.Lm}
\maketitle

\begin{figure}[!hbtp]
\begin{center}
\leavevmode
\includegraphics[width=0.90\columnwidth]{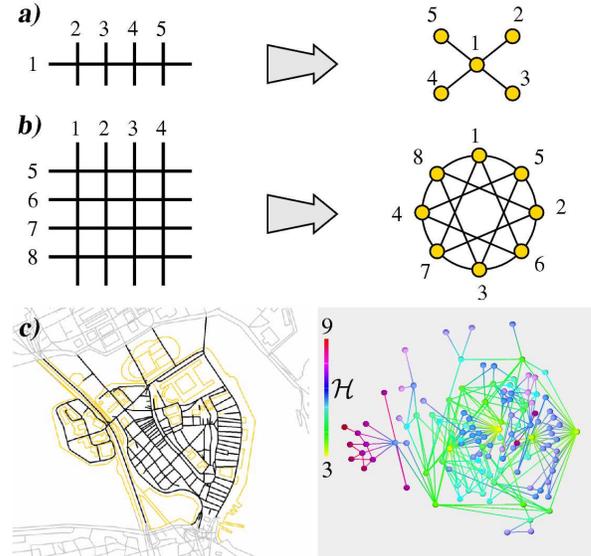}
\end{center}
\vspace{-0.7cm}
\caption{\label{fig1}
Mapping cities (left panels) into information networks (right panels).
In (a) we show how a city 
consisting of one major street and 4
smaller streets maps to a single hub.
This network represents the 
information handling you perform when you orient in such a city.
On the major road 1 you need to know which of 
the exits 2-5 to take
to get to the correct street:
this corresponds to an information of $\log_2(4)=2$ bits.
In (b) we show the city
map of a very planned city,
where each street intersects with many 
perpendicular streets.
For example, it is very easy to go from an north-south (ns)
street to another parallel ns street.
A perpendicular east-west street is first reached with probability $1/4$. 
Next a ns is reached with probability $1/3$, if by assumption a just visited 
street is not visited again. There are four possible paths to the target street 
and therefore one only needs $-\log_2(4 \frac{1}{4} \frac{1}{3}) = \log_2(3)$ 
bits of information to go between 
two parallel streets (see Eq.\ (\ref{sq_city})). 
In (c) we map ``Gamla stan" 
in Stockholm, Sweden, to an information city network. 
Nodes are roughly positioned at the geographical position of 
the corresponding street and color coded according to the
typical amount of information $\mathcal{H}$ needed to locate them.
}
\end{figure}

Traffic and communication between different parts of a complex system
are fundamental elements in maintaining its overall cooperation.
Because a complex system consists of many different parts,
it matters where signals are transmitted.
Thus signaling and traffic is in principle specific, 
with each message going from a unique sender 
to a specific recipient.
One example is living cells, 
where macromolecules are transported between cellular 
components and along micro-tubular highways to 
perform or direct actions on other particular macromolecules \cite{hartwell}. 
This complicated cellular machinery is often
simplified to a molecular network that maps out the 
signaling pathways in the system.
We here will consider a city in a similar perspective,
with communication defined by people that travel from
one specific street to another.
In many cases, the actual traveling distance could easily be 
less restrictive for communication than the 
amount of information needed to locate the correct address.
In this work we will take this perspective to the extreme,
and assume that the travel time/cost of just driving 
along a given road is zero. 
Accordingly we remap a city map to a
dual information representation \cite{jiang}: an information city 
network (Fig.\ \ref{fig1}). Subsequently we will use 
this network to estimate the information needed 
to navigate in a city, and thereby quantify
and compare the complexity of cities.

Imagine that you want to get to a specific street
in the city you are living in.
If you have lived in the city for some time, 
you probably know how to find the street, 
and driving to the destination does not cost any
new information.
However, if you are new in the city, you need travel 
directions along the way to the target.
In this paper we discuss the information value
of such travel directions, or equivalently, we
quantify the information associated to knowing the
city you live in.

Assume that you get your travel directions
in the form of the sequence of roads
that will lead you to the target road. 
These roads form a path of roads with subsequent
intersections. In network language, your trajectory 
can be mapped to a path in an ``information city network'',
where roads map to nodes and 
intersections between roads map to links between the nodes.
This network represents an information 
view of the city, where distances along each road
are effectively set to zero because it
does not demand any information handling to drive between
the crossroads.

In Fig.\ \ref{fig1}(a and b) we present two simple examples
of two caricature cities mapped to such information networks. 
Fig.\ \ref{fig1}(a) shows a particular simple city
consisting of a main road, that together with a collection of smaller
roads define the city. This maps into a single hub,
where all information handling consists in specifying
which of the 4 side roads that is the right one.
In Fig.\ \ref{fig1}(b) we show a slightly more elaborate city,
that resembles modern planned cities.
In that case any street can be accessed from a random 
perpendicular street, and effectively the information 
associated to locate a specific
street is also small.

\begin{figure}[!hbp]
\begin{center}
\leavevmode
\includegraphics[width=1.0\columnwidth]{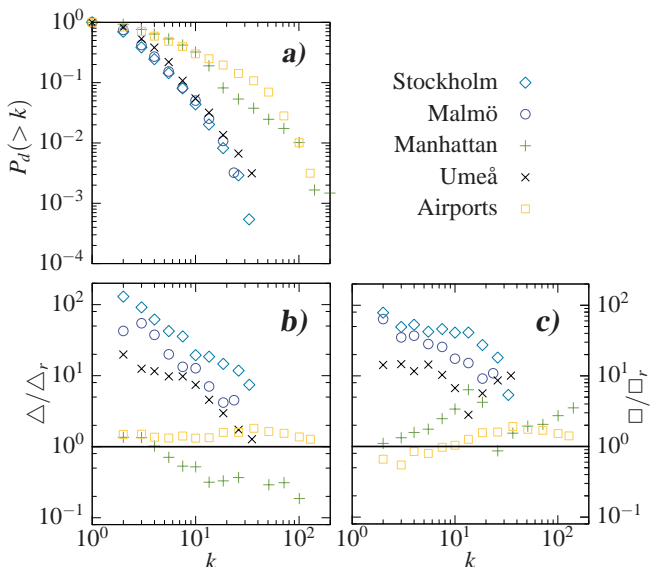}
\end{center}
\vspace{-0.7cm}
\caption{ \label{fig2}
Characterizing traffic networks in terms
of degree distribution (a), and number of 
short loops (b and c).
$P_d(>k)$ is the probability that a node 
has degree $k$ or higher. 
Hence, it is the cumulative degree distribution that is plotted in (a). 
In (b) we show number of loops of length 
3, $\triangle$, that nodes of
degree $k$ participate in, normalized by what this
number of loops would be in a randomized version of
the network, $\triangle_r$, with 100 realizations. The random network is
constructed such that the degree of every node is conserved,
and such that the network remains
globally connected \cite{maslov2002}.
(c) shows the similarly normalized number of loops of length 4, $\Box$.
Both types of loops tend to be over-represented in 
real city networks compared to the randomized ones. 
This reflects locality in the city networks.
}
\end{figure}

In Fig.\ \ref{fig1}(c) we show a part of a real city, ``Gamla stan" 
in Stockholm \cite{teleadress}, Sweden, 
mapped to an information network. 
Long roads with many intersections are 
mapped to major hubs: The network
representation nicely captures that the long roads are 
important for the overall traffic in the system.
For a more systematic study we map a number of different
cities to their information network counterpart,
and examine their basic topological properties 
(Fig.\ \ref{fig2}). For comparison we also show 
another transportation network, consisting
of airports in USA, connected by a link in case
there is a direct 
flight between them \cite{airport}.
In this network, the travel directions are decided in the 
airports and we therefore analyze it with the airports 
as nodes and the flights between airports as links.

For all city networks, and also for the airport network
we observe broad connectivity distributions (Fig.\ \ref{fig2}(a).
However, the local properties differ qualitatively between 
the city networks and the airport network. 
We quantify the locality by the number of small loops
of length 3 (triangles $\triangle$), related to 
clustering \cite{eckmann,newman2003,barabasi}, 
and length 4 (squares $\square$) 
in Fig.\ \ref{fig2}(b and c) normalized 
by their expectation number in random networks with conserved
degree distribution \cite{maslov2002,maslov2002b}. The airport network is 
close to its random counterpart,
whereas the city networks differ substantially
from their random expectations.
The airports are connected with little regards 
to geographical distance, whereas in the cities, 
in particular the short roads
have relatively many loops and thus 
exhibit substantial degree of locality.
Manhattan, selected to represent a planned city, differs 
from the other cities in having few triangles and 
an overabundance of squares associated especially 
to the many streets of connectivity $\sim$$15$ and $\sim$$100$ that, 
respectively, cross the city in east-west and
north-south direction.\\

To characterize the ease or difficulty 
of navigation in different networks, we use the 
``Search Information" $S$ \cite{pnas}.
Imagine a network, in this case an information city network, 
where we start on a node (a street) $s$ and want to locate 
node $t$ (another street)
somewhere else in a connected network with $N$ nodes (streets).
Further, we want to locate $t$ through the
shortest path, or if there are several degenerate shortest paths,
we want to locate $t$ through any of them.
Without prior knowledge, the information needed for locating
a given exit from a node of connectivity $k$, is $\log_2(k)$.
For each path $p(s,t)$ from $s$ to $t$ the probability
to follow it is
\begin{equation}
P[p(s,t)] \; =\; \frac{1}{k_s} \;\; \prod_{j\; \in\; p(s,t)}
\frac{1}{k_j-1},
\end{equation}
with $j$ counting all nodes on the path until the last node before
the target $t$ is reached. The factor $k_j-1$ instead of $k_j$
takes into account the information gained
by following the path, and therefore reducing the number 
of exit links by one.
Thus, the total probability to locate node $t$ 
along any of the degenerate shortest paths is
\begin{equation}
P(s\rightarrow t) = \sum_{\{p(s,t)\}}  P[ p(s,t) ],
\end{equation}
where the sum runs over all degenerate paths that
connect $s$ and $t$.
The total information value of knowing any one
of the degenerate paths between $s$ and $t$ is therefore
\begin{equation}
S(s\rightarrow t) = - \log_2\sum_{\{p(s,t)\}}  P[ p(s,t) ].
\end{equation}
We immediately see that the existence
of many degenerate shortest paths makes it easier to find $t$.
We stress that $S$ should not be confused with  
entropy measures associated to the degree distribution \cite{sole},
or measures related to the dominating eigenvector
of the adjacency matrix \cite{demetrius}.
Instead $S$ is related to specific traffic in the system.

Let us for illustration return to the
``square city" in Fig.\ \ref{fig1}(b),
with $N$ streets divided in $N/2$ north-south ($ns$) streets,
and $N/2$ east-west ($ew$) streets.
Going from any $ns$ street to a particular $ew$ street demands
information about which of the $N/2$ exits we must take. 
This information is $S(ns\rightarrow ew)=\log_2(N/2)$.
On the other hand, if we want to go from one $ns$ 
street to another $ns$ street, we can take anyone of 
the $N/2$ $ew$ streets.
Each path is thus assigned a probability $(2/N)[1/(N/2-1)]$.
But there are in fact $N/2$ degenerate paths, and the 
total information cost for locating parallel 
roads in this square city reduces to
\begin{equation}
\label{sq_city}
S(ns\rightarrow ns)\; =\;
- \log_2( \frac{N}{2} \frac{1}{N/2}
\frac{1}{N/2-1} )
=\log_2(N/2-1),
\end{equation}
reflecting the fact that it does not matter which 
of the $ew$ roads one will use to reach the target road.

\begin{figure}[htbp]
\begin{center}
\leavevmode
\includegraphics[width=0.85\columnwidth]{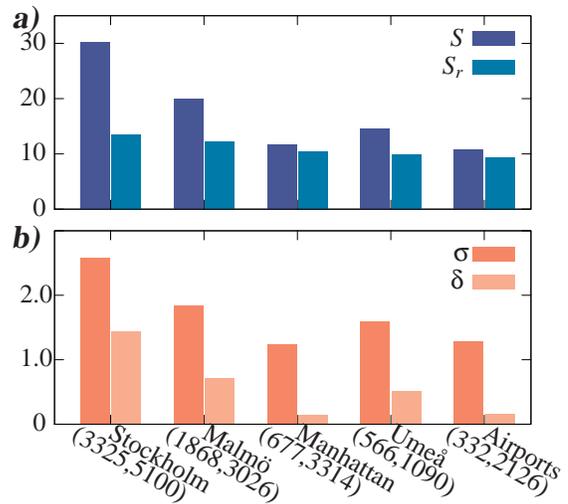}
\end{center}\vspace{-0.7cm}
\caption{ \label{fig3}
(a) shows the average information one 
needs to go from one specific
street to another specific street for some 
city networks \cite{teleadress}, 
and for the network of airports in USA connected 
by commercial airlines \cite{airport}.
($N$,$L$) is, respectively, the number of nodes and links in the networks.
In all cases we compare with the random counterpart of 
the network as described in the caption of Fig.\ \ref{fig2}.
Overall we observe that Manhattan is more 
efficiently organized than the similar sized
Ume{\aa}, but that both are relatively hard 
to navigate in compared with the US airport network.
(b) shows the size-weighted search 
information $\sigma$ together with $\delta$, the corresponding 
difference with the randomized network.
}
\end{figure}

To characterize the overall complexity in finding streets
we calculate the average search information
\begin{equation}
S \; = \; \frac{1}{N^2} \; \sum_{s=1}^N \sum_{t=1}^N S(s,t),
\end{equation}
for a number of cities in Fig.\ \ref{fig3}. 
To evaluate the $S$-values, we for each 
network also calculate the corresponding
$S_r$ for its randomized version.
This random network is constructed such that the
degree of each node is the same as in original network,
and also such that the overall network 
remains connected \cite{maslov2002}.
Thus, comparing $S$ with $S_r$ properly takes into account both
the size of the network, its total number of links as well
as the degree distribution, but not the geometrical constraints. 
The 2-dimensional constraint of a real city is absent in the randomization.
In all cases, including the airline network, we observe that
$S > S_r$. Thus all networks are more difficult to navigate than their
random counterpart (Fig.\ \ref{fig3}(a)).

To take size effects into account we from Eq.\ (\ref{sq_city}) 
expect that $S$ scales as $\log_2(N)$. We therefore define
$\sigma=S/\log_2(N)$ to be able to compare cities of 
different sizes (Fig.\ \ref{fig3}(b)). Furthermore, 
$\delta = (S-S_r)/\log_2(N)$ is interesting, 
since it measures how effectively the city is 
constructed given the length (degree) of the streets (Fig.\ \ref{fig3}(b)).
According to Fig.\ \ref{fig3}(b) Manhattan is relatively easier to 
navigate in than the other cities.
However, neither Manhattan is optimized. 
If Manhattan was constructed as a pure square city (Fig.\ \ref{fig1}(b)) 
the search information would be $S \sim 9$ according to Eq.\ (\ref{sq_city}).

\begin{figure}[htbp]
\begin{center}
\leavevmode
\includegraphics[width=0.95\columnwidth]{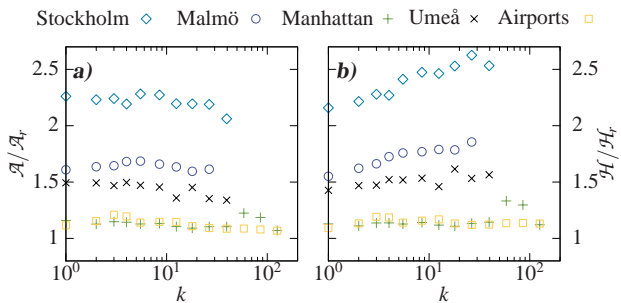}
\end{center}
\vspace{-0.7cm}
\caption{ \label{fig4}%
(a) shows the real-random access ratio 
$\langle \mathcal{A}(k)\rangle /\langle \mathcal{A}_r(k)\rangle $ and 
(b) shows the real-random hide ratio 
$\langle \mathcal{H}(k)\rangle /\langle \mathcal{H}_r(k) \rangle$.
Overall they show similar qualitative behavior. 
Overall (a) and (b) show that the degree of a node plays a minor role
for access $\mathcal{A}$ and hide $\mathcal{H}$.  
}
\end{figure}

To investigate what it is that makes it complicated to navigate in
cities, we in Fig.\ \ref{fig4} measure the information associated to
nodes of different degrees in the network.
We define the access information of a node $s$ by
\begin{equation}
\mathcal{A}_s=\frac{1}{N} \sum_t S(s,t),
\end{equation}
where we sum over all target nodes $t$ in the network.
The quantity $\mathcal{A}_s$ measures the average 
number of questions one needs
to locate a specific street in the network, 
starting from node $s$.
Thus $\mathcal{A}_s$ is a measure
of how good the access to the network is from node $s$.
In Fig.\ \ref{fig4}(a) we show 
$\langle \mathcal{A}(k) \rangle /\langle \mathcal{A}_r(k) \rangle$
averaged over all nodes of degree $k$ versus $k$.
$\langle \mathcal{A}_r(k) \rangle$ is
the average expectation of  $\mathcal{A}(k)$ 
in a randomized network. Note that $\mathcal{H}_t=\frac{1}{N} \sum_s S(s,t) \ne \mathcal{A}_t=\frac{1}{N} \sum_s S(t,s)$. The difference reflects the asymmetry of the endpoints of a path. 
Imagine a small network that consists of a hub with five neighbors. 
The hub is easily reached from any of the neighbors. 
However, starting at the hub it is harder to reach a specific neighbor. 
The hub has low $\mathcal{H}$ and high $\mathcal{A}$ 
and the neighbors have high $\mathcal{H}$ and low $\mathcal{A}$.

The overall feature of Fig.\ \ref{fig4}(a and b) is that
the positioning of the roads with respect to their degree does not explain the 
relatively high values of ${\cal H}$ and ${\cal A}$. However, the degree plays 
another indirect role: The presence of long roads shortens the distances in the 
information network and thereby decreases $S$, especially if degenerate paths exist. 
This is true for Manhattan and the network of airports, 
but not for the three Swedish cities according to their degree distributions (Fig.\ \ref{fig2}(a)).
In the context of city planning, this suggests that for easy navigation 
it is often favorable to replace a big number of shorter streets with 
a few long, provided that they connect remote parts of the network.

When considering 
$\Delta S(l)= \langle S(l) \rangle_{pairs} -
\langle S_r(l) \rangle_{pairs}$ 
as function of distances $l$ between nodes in the city network\cite{foot} 
(not shown), we find that $\Delta S(l)<0$ for distances $l\sim 2$. This 
suggests that local navigation to a neighbor parallel road is 
optimized, whereas the $\Delta S(l)>0$ for $l>3$ reflects a tendency
to protect local neighborhoods by hiding them. Thus the relatively
large $S$ reflects a separation of these neighborhoods.

We also investigated the variance of $\mathcal{A}(k)$ and $\mathcal{H}(k)$
and found that this typically is much larger in real networks, 
compared to their random counterparts. 
This reflects the inhomogeneity in the organization of cities 
(Fig.\ \ref{fig2}(b and c)) with
a fraction of streets being well hidden in remote corners of the cities.
Such corners and local ``islands", 
overrepresented in Stockholm as a consequence of real islands, 
are essentially never present in the random counterparts. 
Many cities are organized hierarchical, where a few main streets connect 
to smaller streets, which in turn connects to even smaller streets.
If a real city was organized purely hierarchical,
with each street connected to
one larger and two smaller streets, then $S=2\:log_2(N)$ for
$N\rightarrow \infty$. In practice this hierarchical organization is
partially broken by intersecting roads (decreasing $S$, e.g.~Manhattan) 
and local neighborhoods or ``islands'' (increasing $S$, e.g.~Stockholm).  
As a consequence $S \approx 2\:log_2(N)$ is only a rough estimate. 
Finally we have measured that locality in the form of 
an excess number of small loops (Fig. \ref{fig2}(b and c)) 
also contributes to $S-S_r$, since small loops introduce 
redundant paths without shortening distances substantially.

We have discussed the organization of 
cities in the perspective of communication 
and presented a way to remap a city map 
to a dual information representation. 
The information representation of a city 
opens for a way to quantify 
the value of knowing it:
A large $S$ means that you have to know a lot 
to find your way around in a city as a newcomer.
In another perspective it is an estimate of the asymmetry 
between traveling a way the first and second time, when 
travel time is included.  
 
We have quantified the intuitive expectation that
Manhattan, and presumably most modern planned 
cities are simple. 
In contrast, historical cities with a complicated past of 
cut and paste construction are more complex.
The observation of a universally large $S$ relatively to $S_r$
in all networks we have investigated means that 
the ability to obtain information is 
relatively more important in these real world networks.
Also it implies that city networks are not optimized for 
communication, as such an optimization would provide
a topology with $S$ even smaller than $S_r$ (Fig. 1b). 
Rather the topologies of real cities, with high $S$, reflect a local
tendency to avoid being exposed to
non-specific traffic.

\end{document}